\documentclass[12pt]{article}
\usepackage{epsfig}
\usepackage{amssymb,amsmath}

\newcommand{\bea}{\begin{eqnarray}}
\newcommand{\eea}{\end{eqnarray}}

 \makeatletter
\def\alt{\mathrel{\mathpalette\gl@align<}}
\def\agt{\mathrel{\mathpalette\gl@align>}}
\def\gl@align#1#2{\lower.6ex\vbox{\baselineskip\z@skip\lineskip\z@
\ialign{$\m@th#1\hfil##\hfil$\crcr#2\crcr\sim\crcr}}} \makeatother
\begin{document}
%\begin{flushright}
%preprint number
%\end{flushright}
%
\vspace*{1.0cm}

\begin{center}
\baselineskip 20pt 
{\Large\bf 
Simple inflationary models \\
in Gauss-Bonnet brane-world cosmology
}
\vspace{1cm}

{\large 
Nobuchika Okada$^{~a}$  and  Satomi Okada
}
\vspace{.5cm}

{\baselineskip 20pt \it
$^a$Department of Physics and Astronomy, University of Alabama, Tuscaloosa, AL35487, USA
} 

\vspace{.5cm}

\vspace{1.5cm} {\bf Abstract}
\end{center}

In light of the recent Planck 2015 results for the measurement of the CMB anisotropy, 
  we study simple inflationary models in the context of the Gauss-Bonnet brane-world cosmology. 
The brane-world cosmological effect modifies the power spectra of scalar and tensor perturbations 
  generated by inflation and causes a dramatic change for the inflationary predictions of 
  the spectral index ($n_s$) and the tensor-to-scalar ratio ($r$) 
  from those obtained in the standard cosmology.  
In particular, the power spectrum of tensor perturbation is suppressed 
  due to the Gauss-Bonnet brane-world cosmological effect, 
  which is in sharp contrast with inflationary scenario in the Randall-Sundrum brane-world cosmology 
  where the power spectrum is enhanced. 
Hence, these two brane-world cosmological scenarios are distinguishable. 
With the dramatic change of the inflationary predictions, the inflationary scenario 
   in the Gauss-Bonnet brane-world cosmology can be tested by more precise measurements 
   of $n_s$ and future observations of the CMB $B$-mode polarization.

\thispagestyle{empty}

%\bigskip
\newpage

\addtocounter{page}{-1}

%%%%%%%%%%%%%%%%%%%%%%%%%%
%\baselineskip 36pt
% Main body
%%%%%%%%%%%%%%%%%%%%%%%%%%
\baselineskip 18pt
%%%%%%%%%%%%%%%%%%%%%%%%%%

%%%%%%%%%%%%%%%%%%%%%%%%
\section{Introduction} 
%%%%%%%%%%%%%%%%%%%%%%%%
Inflationary universe is the standard paradigm in the modern cosmology~\cite{inflation1, inflation2, chaotic_inflation, inflation4}, 
  by which serious problems of the standard big-bang cosmology, 
  such as the flatness and horizon problems, can be solved. 
In addition,  inflationary universe provides the primordial density fluctuations 
  as seeds for the formation of the large scale structure observed in the present universe.  
Various inflationary models have been proposed with typical inflationary predictions 
  for the spectral index ($n_s$), the tensor-to-scalar ratio ($r$), the running of the spectral index ($\alpha=d n_s/d \ln k$), 
 and non-Gaussianity of the primordial perturbations. 
These predictions are currently tested by precise measurements of the cosmic microwave background (CMB) anisotropy 
   by the Wilkinson Microwave Anisotropy Probe (WMAP)~\cite{WMAP9} and the Planck satellite~\cite{Planck2013} experiments. 
Future cosmological observations are expected to become more precise towards discriminating inflationary models.

Very recently, the Planck collaboration has updated their results from Planck 2013 results, 
  and provided the more stringent constraints on the inflationary predictions. 
Motivated by this Planck 2015 results~\cite{Planck2015}, 
   we study inflationary scenario in the context of the brane-world cosmology.  
The brane-world cosmology is based on the so-called RS II model first proposed 
  by Randall and Sundrum (RS)~\cite{RS2}, 
  where the Standard Model particles are confined on a "3-brane"  at a boundary embedded 
  in 5-dimensional anti-de Sitter (AdS) space-time. 
Because of the AdS space-time geometry, massless graviton in 4-dimensional effective theory 
  is localized around the brane on which the Standard Model particles reside, 
  while Kaluza-Klein gravitons are delocalized toward infinity. 
As a result, the 4-dimensional Einstein-Hilbert action is reproduced at low energies. 
A realistic cosmological solution in the RS II setup has been found in \cite{RS2solution}, 
  which leads to the Friedmann equation in the 4-dimensional standard cosmology 
  at low energies, while a non-standard expansion law at high energies.
Since then, the RS II cosmology has been intensively studied~\cite{braneworld}. 
The non-standard evolution of the early universe causes modifications of  
  a variety of phenomena in particle cosmology, such as the dark matter relic abundance~\cite{DM_BC}, 
  baryogensis via leptogenesis~\cite{LG_BC}, and gravitino productions in the early universe~\cite{gravitino_BC}. 
A chaotic inflation with a quadratic inflaton potential has been examined in \cite{inflation_BC}, 
  and it has been shown that the inflationary predictions are modified from those in the 4-dimensional standard cosmology. 
In particular, the power spectrum of tensor perturbation is found to be enhanced 
   in the presence of the 5-dimensional bulk ~\cite{PT_BC}. 
Taking the brane-world cosmological effect into account, the textbook chaotic inflation models 
   with the quadratic and quartic potentials (monomial potentials in general) 
   have been analyzed in \cite{inflation_models_BC, inflation_models_BC2}.  
In light of the observation of CMB $B$-mode polarization reported 
  by the Background Imaging of Cosmic Extragalactic Polarization (BICEP2) collaboration~\cite{BICEP2}, 
  the Higgs potential and the Coleman-Weinberg potential models, 
  in addition to the textbook chaotic inflation models, have been analyzed in \cite{OO2}. 
It has been shown that these simple inflationary models except the quartic potential model 
   can nicely fit the BICEP2 result with the enhancement of the tensor-to-scalar ratio 
   due to the brane-world cosmological effects. 
Unfortunately, recent joint analysis of BICEP2/Keck Array and Planck data~\cite{BKP} has concluded 
  that uncertainty of dust polarization dominates the excess observed by the BICEP2 experiment.

In this paper, we investigate the simple inflationary models in the Gauss-Bonnet (GB) brane-world cosmology, 
  where the RS II model  is extended by adding the Gauss-Bonnet invariant~\cite{GB_invariant}. 
The Friedmann equation for the GB brane-world cosmology has been found in \cite{GB}, 
  with which we analyze the inflationary models and compare the inflationary predictions 
  with the Planck 2015 results. 
For previous work with simple monomial inflaton potentials in the GB brane-world cosmology, 
  see \cite{TSM}.  
See also \cite{IPN} for similar discussion in light of the BICEP2 result. 
At high energies, where the GB invariant dominates the evolution of the universe (GB regime), 
  the expansion law is quite different from that in the RS II cosmology.\footnote{
See \cite{DM_GB} for the modification of dark matter physics in the GB brane-world cosmology. 
}
Furthermore, it has been found that in the GB regime, the power spectrum of tensor perturbation 
   is suppressed compared to that in the 4-dimensional standard cosmology~\cite{PT_GB}. 
Therefore, the inflationary predictions in the GB brane-world cosmology are altered from 
   those in the standard cosmology as well as the RS brane-world cosmology.

In the next section, we briefly review the GB cosmological model and give the Friedmann equation 
   in the GB regime. 
In Sec.~3, we analyze simple inflationary models based on  
  the textbook models, the Higgs potential and the Coleman-Weinberg potential models. 
We obtain the inflationary predictions as a function of a parameter ($\mu$) 
  characterizing the GB brane-world cosmological effect and compare the predictions 
  with the Planck 2015 results. 
We also show responses of the parameters in the inflationary models to the parameter $\mu$. 
The last section is devoted to conclusions.

%%%%%%%%%%%%%%%%%%%%%%%%%%%%%%%%%%%%%%%%%%%%%%%%%%%%%%%%%%%
\section{The GB brane-world cosmology}
%%%%%%%%%%%%%%%%%%%%%%%%%%%%%%%%%%%%%%%%%%%%%%%%%%%%%%%%%%%

Motivated by string theory considerations, it would be natural to extend the RS II cosmological model 
  by adding higher curvature terms~\cite{GB_invariant}.  
Among a variety of such terms, the GB invariant is of particular interests in five dimensions, 
 since it is a unique nonlinear term in curvature which yields second order gravitational field equations. 
The extended RS II action with the GB invariant is given by 
\bea
 {\cal S} &=& \frac{1}{2\kappa_5^2} \int d^5x 
 \sqrt{-g_5}
 \left[
 - 2 \Lambda_5+ {\cal R} + 
 \alpha \left( {\cal R}^2 -4 {\cal R}_{ab} {\cal R}^{ab} 
 + {\cal R}_{a b c d}{\cal R}^{a b c d} \right) \right] \nonumber \\
&-& \int_{brane} d^4x 
 \sqrt{-g_4} \left( m_\sigma^4 + {\cal L}_{matter} \right), 
\label{GBaction}
\eea
where $\kappa_5^2=8\pi/M_5^3$ with the five-dimensional Planck mass $M_5$, $ m_\sigma^4 >0 $ is a brane tension, 
  and $ \Lambda_5 <0 $ is the bulk cosmological constant. 
The limit $\alpha \to 0$ recovers the RS II model.

Imposing a $Z_2$ parity across the brane in an anti-de Sitter bulk 
  and modeling the matters on the brane as a perfect fluid, 
  the Friedmann equation on the spatially flat brane has been found to be \cite{GB}
\bea 
 \kappa_5^2(\rho + m_\sigma^4) = 
  2 \mu \sqrt{1+\frac{H^2}{\mu^2}}
  \left( 3 - \beta +2 \beta \frac{H^2}{\mu^2} \right) ,  
\label{GBFriedmann1} 
\eea
where $\beta = 4 \alpha \mu^2 = 1-\sqrt{1 + 4 \alpha \Lambda_5/3}$.  
The model has four free parameters, $\kappa_5$, $m_\sigma$, $\mu$ and $\beta$, 
  which are constrained by phenomenological requirements as follows. 
To reproduce the Friedmann equation of the standard cosmology with a  vanishing cosmological constant 
   for the limit $H^2/\mu^2 \ll 1$,  we have two conditions, 
\bea 
 \kappa_5^2 m_\sigma^4 = 2 \mu (3-\beta), \; \;  \kappa_4^2 = 1/M_{P}^{2} 
  = \frac{\mu}{1+\beta} \kappa_5^2 , 
\label{relations} 
\eea
where $M_{P}=M_{Pl}/\sqrt{8 \pi}$ is the reduced Planck mass with $M_{Pl}=1.22 \times 10^{19}$ GeV. 
From now on, we use the Planck unit, $M_P=1$. 
The modified Friedmann equation can be rewritten in the useful form \cite{GBsolution3} 
\bea
&& H^2 = \frac{\mu^2}{\beta} 
 \left[ (1- \beta) \cosh \left(\frac{2 \chi}{3} \right) -1 \right], 
  \nonumber \\ 
&& \rho + m_\sigma^4 = m_\alpha^4 \sinh \chi , 
\label{GBFriedmann2} 
\eea
where $\chi$ is a dimensionless measure of the energy density and 
\bea
 m_\alpha^4 = \sqrt{ 
 \frac{8 \mu^2 (1-\beta)^3}{\beta \kappa_5^4}} 
 = 2 \mu^2   \sqrt{2 \frac{(1-\beta)^3}{\beta (1+\beta)^2}}. 
\eea
Here we have used Eq.~(\ref{relations}) to eliminate $\kappa_5$ in the last equality. 
In the same way, we express $m_\sigma$ as 
\bea
 m_\sigma^4 = 2 \mu^2 \left( \frac{3-\beta}{1+\beta} \right). 
\eea

%%%%%%%%%%%%%%%%%%%%%%%%%%%%%%%%%%%%%%%%%%%%
\begin{figure}[htbp]
\begin{center}
\includegraphics[width=0.45\textwidth,angle=0,scale=1.02]{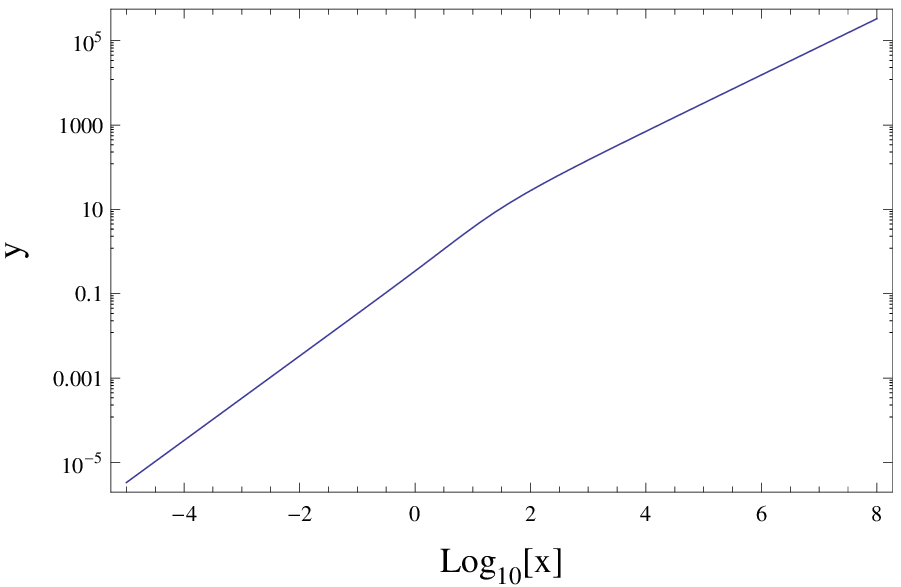} \hspace{0.5cm}
\includegraphics[width=0.45\textwidth,angle=0,scale=1.02]{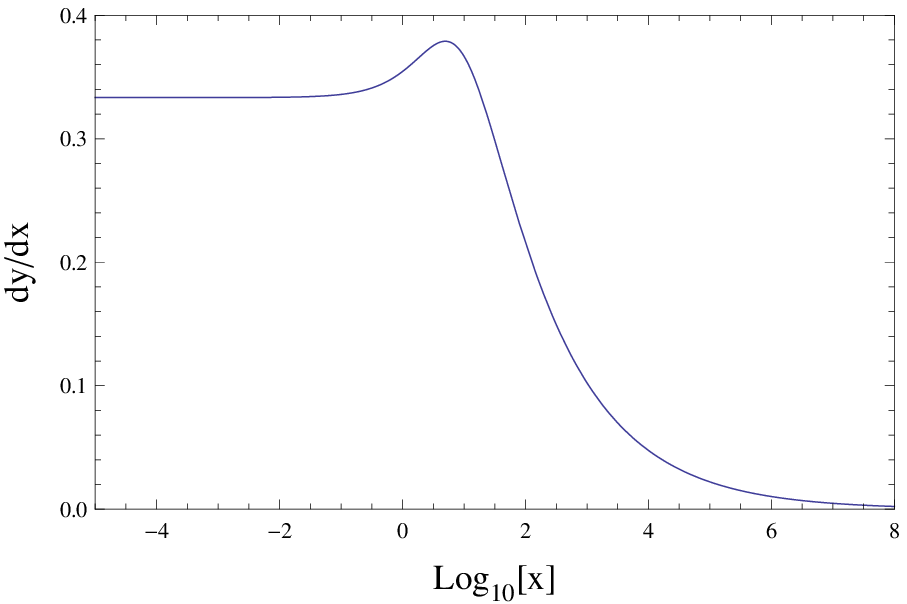}
\end{center}
\caption{
In the Planck unit ($M_P=1$), $y=H^2/\mu^2$ as a function of $x=\rho/\mu^2$ (left), 
  and the first derivative of $y$ with respect to $x$ (right).  
}
\label{fig:1}
\end{figure}
%%%%%%%%%%%%%%%%%%%%%%%%%%%%%%%%%%%%%%%%%%%% 

The evolution of the GB brane-world cosmology is characterized 
   by the two mass scales, $m_\alpha$ and $m_\sigma$. 
Expanding Eq.~(\ref{GBFriedmann2}) with respect to $\chi$, we find three regimes for $m_\alpha > m_\sigma$. 
The GB regime for $\rho \gg m_\alpha^4$, 
\bea 
 H^2 \simeq 
 \left(  \frac{1+\beta}{4 \beta} \mu  \rho 
  \right)^{2/3},   
\eea
the RS regime for $m_\alpha^4 \gg \rho \gg m_\sigma^4$, 
\bea 
 H^2 \simeq 
  \frac{\rho^2}{6 m_\sigma^4} ,
\eea
and the standard regime for $m_\sigma^4 \gg \rho $, 
\bea 
 H^2 \simeq \frac{\rho}{3} .
\eea
Since we are interested in the GB regime, let us simplify the evolution of the universe 
  by imposing the condition $m_\alpha = m_\sigma$, which leads to 
\bea
 3 \beta^3 -12 \beta^2 + 15 \beta -2 = 0   
\eea
 and hence, $\beta=0.151$. 
In this case, the RS regime is collapsed, and there are only two regimes in the evolution of the universe.  
Now we rewrite the modified Friedmann equation as 
\bea 
 (1+\beta) \frac{\rho}{\mu^2}  + 2 (3 -\beta )
=   2 \sqrt{1+\frac{H^2}{\mu^2}}
  \left( 3 - \beta +2 \beta \; \frac{H^2}{\mu^2} \right) ,  
\label{GBFriedmann} 
\eea
which is characterized by only one free parameter $\mu$. 
Solving this equation, we obtain $H^2/\mu^2$ as a function of $\rho/\mu^2$. 
In Fig.~\ref{fig:1}, we plot $y=H^2/\mu^2$ as a function of $x=\rho/\mu^2$ (left) and 
  the first derivative of $y$ with respect to $x$ (right).  
We can see that $y \propto x^{2/3}$ for $x \gg 1$ (GB regime), while $y \propto x$ for $x \ll 1$ (standard regime).

%%%%%%%%%%%%%%%%%%%%%%%%%%%%%%%%%%%%%%%%%%%%%%%%%%%%%%%%%%%
\section{Simple inflationary models in GB brane-world cosmology}
%%%%%%%%%%%%%%%%%%%%%%%%%%%%%%%%%%%%%%%%%%%%%%%%%%%%%%%%%%%
We first give basic formulas used in the following analysis for inflationary models. 
In the slow-roll inflation, the Hubble parameter is given by the solution of Eq.~(\ref{GBFriedmann}) 
   with $ \rho = V(\phi)$, where $V$ is a potential of the inflaton field $\phi$. 
Since the inflaton is confined on the brane, the power spectrum of scalar perturbation obeys  
  the same formula as in the standard cosmology, except for the modification of the Hubble parameter~\cite{inflation_BC}, 
\begin{eqnarray}
   {\cal P}_{\cal S} =\frac{9}{4 \pi^2} \frac{H^6}{(V')^2}, 
\end{eqnarray}
 where the prime denotes the derivative with respect to the inflaton field $\phi$. 
For the pivot scale chosen at $k_0=0.002$ Mpc$^{-1}$, the power spectrum of scalar perturbation 
  is constrained as $ {\cal P}_{\cal S}(k_0) = 2.196 \times 10^{-9}$ by the Planck 2015 results~\cite{Planck2015}. 
By using  the slow-roll parameters defined as  
\begin{eqnarray}
\epsilon = \frac{V^\prime}{6 H^2} \left( \ln H^2 \right)^\prime, \; \; \eta =\frac{V^{\prime \prime}}{3 H^2} , 
\end{eqnarray}
the spectral index is given by
\begin{eqnarray}
 n_s -1 = \frac{d \ln {\cal P}_{\cal S}}{d\ln k} =-6 \epsilon + 2 \eta .
\end{eqnarray}
Hence, the running of the spectral index, $\alpha \equiv dn_s/d\ln k$,  is given by 
\begin{eqnarray}
  \alpha = \frac{dn_s}{d\ln k} =\frac{V^\prime}{3 H^2}  \left( 6 \epsilon^\prime - 2 \eta^\prime   \right).  
\end{eqnarray}
On the other hand, in the presence of the extra dimension where graviton resides, 
  the power spectrum of tensor perturbation is modified to be~\cite{PT_GB} 
\begin{eqnarray}
 {\cal P}_{\cal T} =8  \left(  \frac{H}{2 \pi} \right)^2 F(x_0)^2, 
 \label{P_T}
\end{eqnarray}
where  $x_0 = \sqrt{H^2/\mu^2}$, and 
\begin{eqnarray}
 F(x)= \left( \sqrt{1+x^2} - \left(  \frac{1-\beta}{1+\beta} \right)  x^2 \ln\left[ \frac{1}{x}+\sqrt{1+\frac{1}{x^2}} \right]   \right)^{-1/2}. 
\end{eqnarray} 
For $x \ll 1$, $F(x)^2 \simeq 1$, and Eq.~(\ref{P_T}) reduces to the formula in the standard cosmology. 
For $x_0 \gg 1$, $F(x)^2 \simeq \frac{1+\beta}{2 \beta x} \simeq 3.81/x $, 
  so that the power spectrum of tensor perturbation is suppressed in the GB regime. 
Note that in the limit $\beta =0$, the above $F(x)$ is reduced to the one found for the RS brane-world cosmology in \cite{PT_BC}. 
In the RS case ($\beta=0$), $F(x)^2 \simeq 1.5 x $ for $x \gg 1$, 
  and the power spectrum of tensor perturbation is enhanced. 
The tensor-to-scalar ratio is defined as $r =  {\cal P}_{\cal T}/ {\cal P}_{\cal S}$.

The e-folding number is given by 
\begin{eqnarray}
N_0 = \int_{\phi_e}^{\phi_0} d\phi \; \frac{3 H^2}{V^\prime} ,
\end{eqnarray}
where $\phi_0$ is the inflaton VEV at horizon exit of the scale corresponding to $k_0$, 
  and $\phi_e$ is the inflaton VEV at the end of inflation, which is defined by ${\rm max}[\epsilon(\phi_e), | \eta(\phi_e)| ]=1$.
In the standard cosmology, we usually consider $N_0=50-60$ in order to solve the horizon problem. 
Since the expansion rate in the GB brane-world cosmology is smaller than the standard cosmology case, 
  we may expect a smaller value of the e-folding number. 
However, since the e-folding number also depends on reheating temperature after inflation, 
  we consider $N_0=50$ and $60$ as reference values, as usual in the standard cosmology.

%%%%%%%%%%%%%%%%%%%%%%%%%%%
\subsection{Textbook inflationary models}
%%%%%%%%%%%%%%%%%%%%%%%%%%% 
Let us first consider the textbook chaotic inflation model with a quadratic potential~\cite{chaotic_inflation},  
\begin{eqnarray}
V =\frac{1}{2} m^2 \phi^2. 
\end{eqnarray}
In the standard cosmology, simple calculations lead to the following inflationary predictions:
\begin{eqnarray}
n_s=1-\frac{4}{2 N_0+1},  \; \; 
r=\frac{16}{2 N_0+1}, \; \; 
\alpha= - \frac{8}{(2 N_0+1)^2}. 
\end{eqnarray}
The inflaton mass is determined so as to satisfy the power spectrum measured by the Planck satellite experiment, 
 ${\cal P}_{\cal S}(k_0)=2.196 \times 10^{-9}$:
\begin{eqnarray}
 m [{\rm GeV}]= 1.45 \times 10^{13}  \left(  \frac{121}{2 N_0+1}\right). 
\end{eqnarray}

In the GB brane-world cosmology, these inflationary predictions in the standard cosmology 
 are altered due to the modified Friedmann equation in Eq.~(\ref{GBFriedmann}).
In the limit of $H/\mu \gg 1$,  the Hubble parameter is simplified as 
\bea 
  H^2 \simeq \left( \frac{1+\beta}{4 \beta} \mu  V(\phi)    \right)^{2/3} .  
\eea 
Using this expression, we can easily find the inflationary predictions as 
\begin{eqnarray}
n_s=1-\frac{6}{4 N_0+3},  \; \; 
r=\frac{32}{4 N_0+3}, \; \; 
\alpha= - \frac{24}{(4 N _0+3)^2}. 
\label{phi2_GBlimit}
\end{eqnarray} 
 From the experimental constraint ${\cal P}_{\cal S}(k_0)=2.196 \times 10^{-9}$, 
we also find (in the Planck unit)
\bea 
  m \mu \simeq 7.20 \times 10^{-11} \left(  \frac{243}{4 N_0+3} \right)^{3/2}
\eea
Since $\mu \propto M_5^3$ (see Eq.~(\ref{relations})), the inflaton mass ($m$) becomes larger, 
  as the 5-dimensional Planck mass is lowered, 
This is in sharp contrast with a relation found in the RS brane-world cosmology 
 (see, for example, Eq.~(16) in \cite{OO2}), 
\begin{eqnarray}
 \frac{m}{M_5} \simeq 1.26 \times 10^{-4} \left(  \frac{121}{2 N_0+1}\right)^{5/6} ,
 \label{m/M5}
\end{eqnarray} 
where the inflaton mass becomes smaller as $M_5$ is lowered.

%%%%%%%%%%%%%%%%%%%%%%%%%%%%%%%%%%%%%%%%%%%%
\begin{figure}[htbp]
\begin{center}
\includegraphics[width=0.45\textwidth,angle=0,scale=1.04]{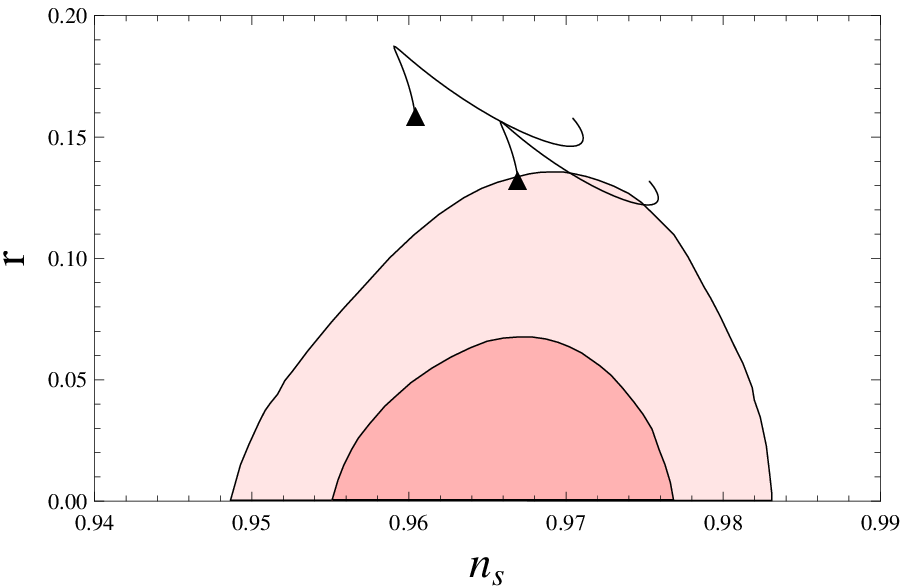} \hspace{0.5cm}
\includegraphics[width=0.45\textwidth,angle=0,scale=1.08]{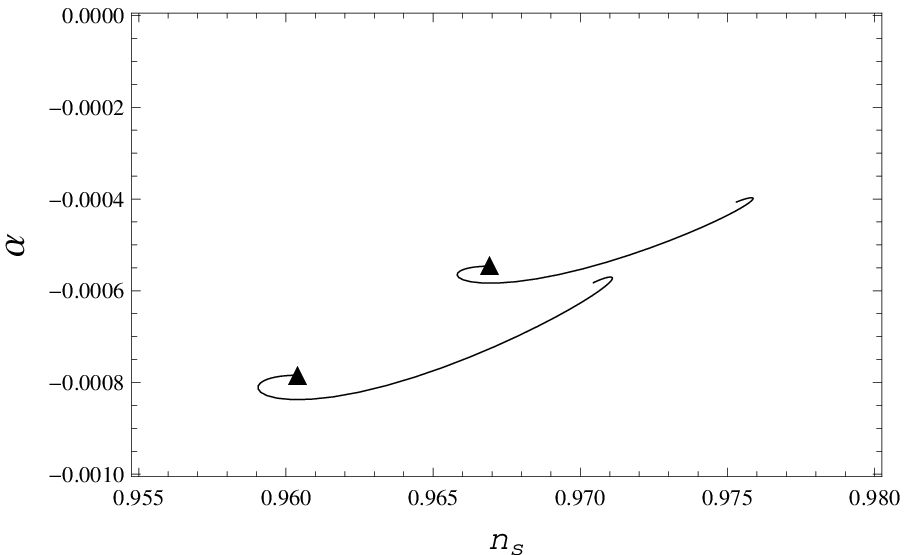}
\end{center}
\caption{
The inflationary predictions for the quadratic potential model: $n_s$ vs. $r$ (left panel) and $n_s$ vs. $\alpha$ (right panel)
  for various $\mu$ values with $N_0=50$ and $60$ from left to right. 
The contours on the background are the 68\% and the 95\% C.L. from the Planck 2015 ({\it Planck} TT+lowP)~\cite{Planck2015}. 
The black triangles are the predictions of the textbook quadratic potential model in the standard cosmology, 
  which are reproduced for $H/\mu \ll 1$. 
As $\mu$ is lowered, the inflationary predictions approach the values in Eq.~(\ref{phi2_GBlimit}). 
In each line, the turning points appear according to a non-trivial behavior of 
  the first derivative of the Hubble parameter shown in the left panel of Fig.~\ref{fig:1}. 
For $N_0=60$, the inflationary predictions lie inside the contour at 95\% C.L. 
  for $1.89 \times 10^{12} \leq \mu[{\rm GeV}] \leq 3.44 \times 10^{12}$. 
}
\label{fig:phi2}
\end{figure}
%%%%%%%%%%%%%%%%%%%%%%%%%%%%%%%%%%%%%%%%%%%% 

%%%%%%%%%%%%%%%%%%%%%%%%%%%%%%%%%%%%%%%%%%%%
\begin{figure}[htbp]
\begin{center}
\includegraphics[width=0.45\textwidth,angle=0,scale=1.05]{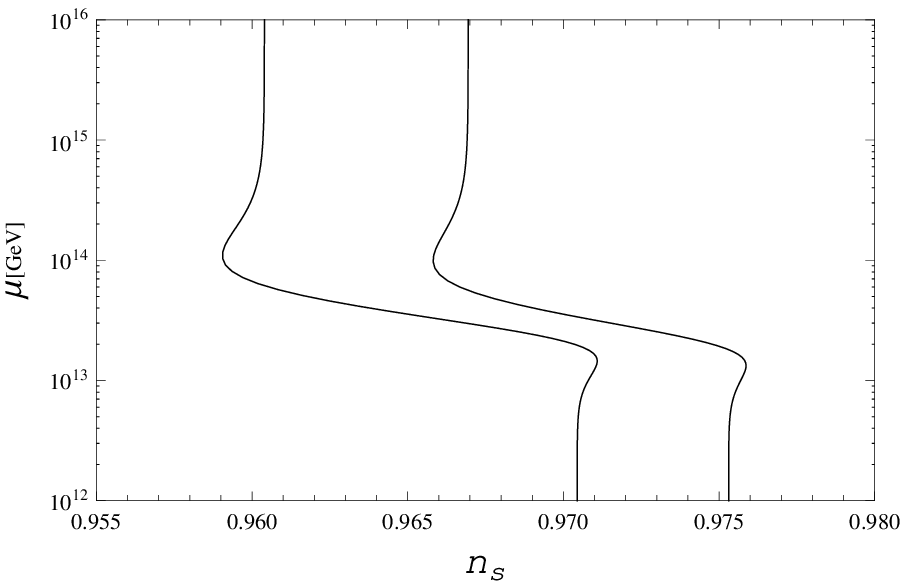} \hspace{0.5cm}
\includegraphics[width=0.45\textwidth,angle=0,scale=1.05]{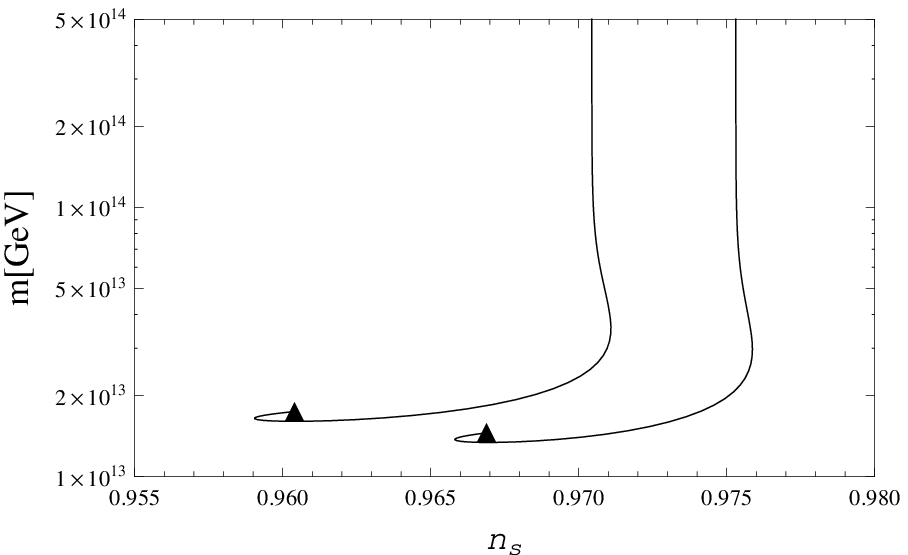}
\end{center}
\caption{
Relations between $n_s$ and $\mu$ (left panel) and between  $n_s$ and $m$ (right panel), 
  for $N_0=50$ and $60$ from left to right. 
The black triangles denote the predictions in the standard cosmology.
}
\label{fig:phi2_mass}
\end{figure}
%%%%%%%%%%%%%%%%%%%%%%%%%%%%%%%%%%%%%%%%%%%% 

We calculate the inflationary predictions for various values of $\mu$ with fixed e-folding numbers, 
  and show the results in Fig.~\ref{fig:phi2}. 
In the left panel, the inflationary predictions for $N_0=50$ and $60$ from left to right are shown, 
 along with the contours (at the confidence levels of 68\% and 95\%) 
 from the Planck 2015 results ({\it Planck} TT+lowP)~\cite{Planck2015}.  
The black triangles represent the predictions of the quadratic potential model in the standard cosmology. 
As $\mu$ is lowered, the inflationary predictions approach the values in Eq.~(\ref{phi2_GBlimit}). 
In each line, some turning points appear according to a non-trivial behavior of 
  the first derivative of the Hubble parameter shown in the right panel of Fig.~\ref{fig:1}.
The quadratic potential model in the standard cosmology is not favored by the Planck 2015 results. 
Although we have found that for $1.89 \times 10^{12} \leq \mu[{\rm GeV}] \leq 3.44 \times 10^{12}$, 
  the inflationary predictions can be consisted 
  with the Planck results at 95\% C.L., the GB brane-world effect cannot significantly improve the fit. 
The results for the running of the spectral index ($n_s$ vs. $\alpha$) is shown in the right panel, 
  for $N_0=50$ and $60$ from left to right. 
 As usual for simple inflationary models, the predicted $|\alpha|$ is vey small and 
  consistent with the Planck 2015 results~\cite{Planck2015}, $\alpha=-0.0126^{+0.0098}_{-0.0087}$ ({\it Planck} TT+lowP).  
We also show our results for the relations between  $n_s$ vs. $\mu$, 
  and $n_s$ vs. $m$ in Fig.~\ref{fig:phi2_mass}. 
Comparing two panels in Fig.~\ref{fig:phi2_mass}, we see that the inflaton mass is increased as $\mu$ is lowered.

Next we analyze the textbook quartic potential model, 
\begin{eqnarray}
V =\frac{\lambda}{4!} \phi^4. 
\end{eqnarray}
In the standard cosmology, we find the following inflationary predictions:
\begin{eqnarray}
n_s=1-\frac{6}{2 N_0+3},  \; \; 
r=\frac{32}{2 N_0+3}, \; \; 
\alpha= - \frac{12}{(2 N_0+3)^2}. 
\end{eqnarray}
The quartic coupling ($\lambda$) is determined by the power spectrum measured by the Planck satellite experiment, 
 ${\cal P}_{\cal S}(k_0)=2.196 \times 10^{-9}$ at the pivot scale $k_0=0.002$ Mpc$^{-1}$,  as 
\begin{eqnarray}
  \lambda = 8.39 \times 10^{-13}  \left(  \frac{123}{2 N_0+3}\right)^3. 
\end{eqnarray}

\begin{figure}[htbp]
\begin{center}
\includegraphics[width=0.45\textwidth,angle=0,scale=1]{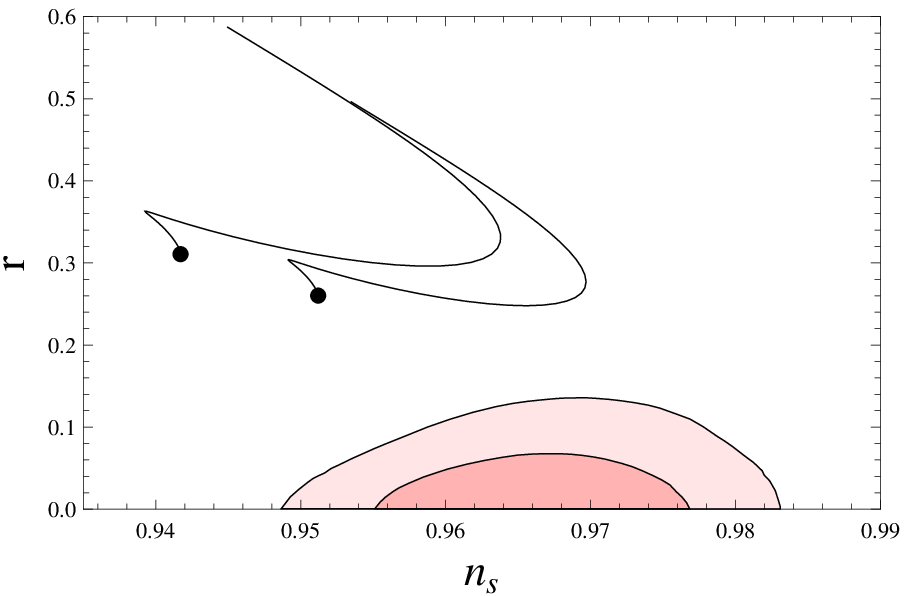} \hspace{0.5cm}
\includegraphics[width=0.45\textwidth,angle=0,scale=1.05]{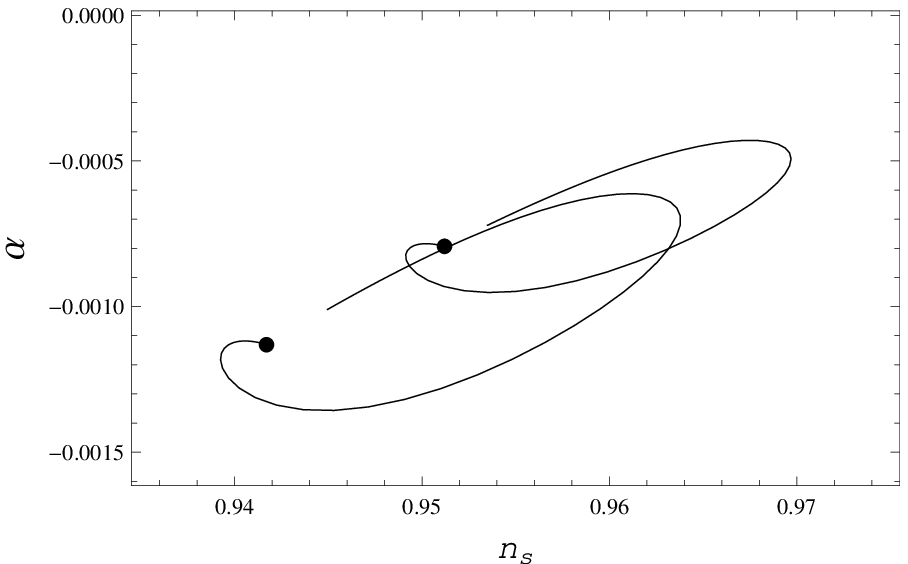}
\end{center}
\caption{
The inflationary predictions for the quartic potential model: $n_s$ vs. $r$ (left panel) and $n_s$ vs. $\alpha$ (right panel)
  with various $\mu$ values for $N_0=50$ and $60$ from left to right. 
The contours on the background are the same as in Fig.~\ref{fig:phi2}.
The black points are the predictions of the textbook quartic potential model in the standard cosmology. 
The inflationary predictions are moving anti-clockwise along the contours as $\mu$ is lowered. 
}
\label{fig:phi4}
\end{figure}

We calculate the inflationary predictions for various values of $\mu$ with fixed e-folding numbers $N_0=50$ and $60$. 
Our results are shown in Fig.~\ref{fig:phi4}. 
In the left panel, the inflationary predictions for $N_0=50$ and $60$ from left to right are shown, 
 along with the contours (at the confidence levels of 68\% and 95\%) from 
   the Planck 2015 results, as in Fig.~\ref{fig:phi2}. 
The results for the running of the spectral index ($n_s$ vs. $\alpha$) is shown in the right panel, 
 for $N_0=$50 and $60$ from left to right. 
 The black points represent the predictions in the standard cosmology presented above. 
In Fig.~\ref{fig:phi4}, the inflationary predictions are moving anti-clockwise along the contours as $\mu$ is lowered. 
The quartic potential model is disfavored by the Planck 2015 results, 
  and no improvement for the fit is obtained by the GB brane-world cosmological effect. 
  
%%%%%%%%%%%%%%%%%%%%%%%%%%%%%%%%%%%%%%%%%%%%
%%%%%%%%%%%%%%%%%%%%%%%%%%%%%%%%%%%%%%%%%%%% 

%%%%%%%%%%%%%%%%%%%%%%%%%%%%%%%%%%%%%%%%%%%%%%%%%%%%%
\subsection{Higgs potential model}
%%%%%%%%%%%%%%%%%%%%%%%%%%%%%%%%%%%%%%%%%%%%%%%%%%%%%
The next simple inflationary model we will consider is based on the Higgs potential of the form~\cite{Higgs_potential_model}
\begin{eqnarray}
V= \lambda \left( \phi^2 -v^2 \right)^2, 
\end{eqnarray}
where $\lambda$ is a real, positive coupling constant, $v$ is a VEV of the inflaton $\phi$. 
Here, we assume that the inflaton is a real scalar for simplicity, but it is easy to extend 
   the present model to the Higgs model where the inflaton field breaks a gauge symmetry by its VEV. 
We refer, for example, Refs.~\cite{hp_corrections, BL_inflation} for recent discussion 
  about such a class of inflationary models, where quantum corrections of the Higgs potential 
  are also taken into account.

For analysis of this inflationary scenario, we can, in general, consider two cases for the initial inflaton VEVs, 
  namely, (i) $\phi_0 > v$ and (ii) $\phi_0 < v$. 
However,  in this paper, we concentrate on the case (ii), since the results for the case (i) are largely 
  covered by those in the previous subsection.  
To see this, we rewrite the potential in terms of a new inflaton field $\chi$ defined as $\phi=\chi +v$, 
\begin{eqnarray}
V= \lambda  \left( 4 v^2 \chi^2 +4 v \chi^3 +\chi^4 \right). 
\end{eqnarray}
Clearly, when an initial value of inflaton ($\chi_0$) satisfies a condition, $\chi_0/v \ll 1$, 
  the inflaton potential is dominated by the quadratic term. 
Hence, the inflationary predictions are similar to those for the textbook quadratic potential model. 
On the other hand,  when a condition of $\chi_0/v \gg 1$ is satisfied, the quartic term dominates 
  the inflaton potential, and the inflationary predictions in this case are covered by 
  the analysis for the textbook quartic potential model. 
Therefore, the inflationary predictions of the model in the case (i) interpolate the inflationary predictions 
  of the textbook quadratic and quartic potential models by varying the inflaton VEV from $v=0$ to $v \gg 1$.

%%%%%%%%%%%%%%%%%%%%%%%%%%%%%%%%%%%%%%%%%%%%
\begin{figure}[htbp]
\begin{center}
\includegraphics[width=0.45\textwidth,angle=0,scale=1.0]{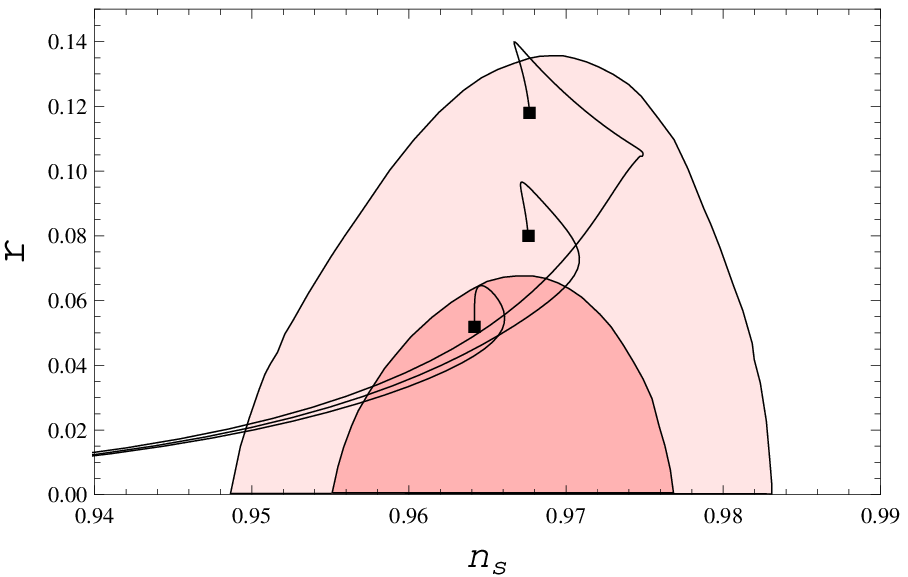} \hspace{0.5cm}
\includegraphics[width=0.45\textwidth,angle=0,scale=1.05]{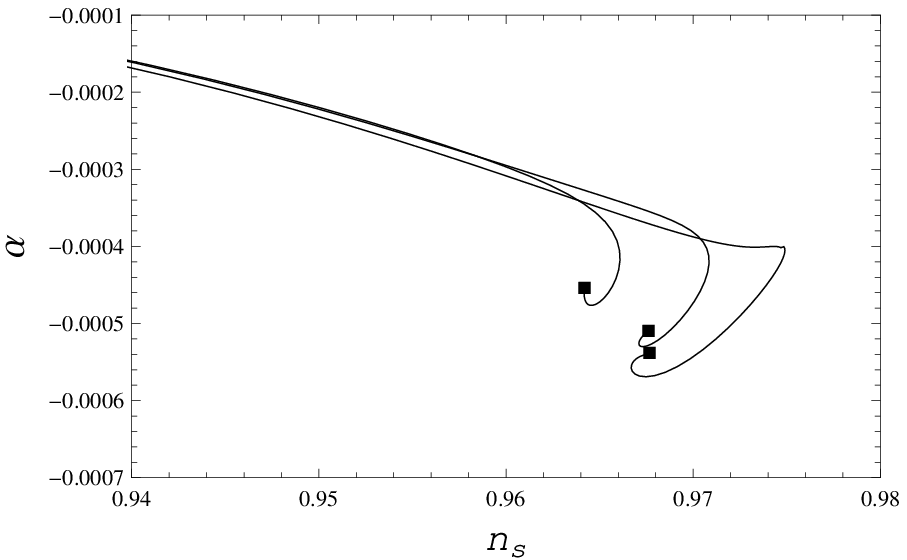}
\end{center}
\caption{
The inflationary predictions for the Higgs potential model: $n_s$ vs. $r$ (left panel) and $n_s$ vs. $\alpha$ (right panel)
  for various $\mu$ values with fixed $v=20$, $30$ and $100$ from left to right. 
The  contours on the background are the same as in Fig.~\ref{fig:phi2}. 
Here we have fixed the number of e-foldings $N_0=60$. 
The black squares from bottom to top (top to bottom) in the left panel (the right panel) 
  denotes the inflationary predictions in the standard cosmology limit for $v=20$, $30$ and $100$, respectively. 
As $\mu$ is lowered, the inflationary predictions are deviating from the predictions in the standard cosmology. 
We have found the lower bounds on  $\mu[{\rm GeV}]/10^{12} \geq 5.13$, $3.41$ and $1.03$ 
  for $v=20$, $30$ and $100$, respectively, for the inflationary predictions 
  to lie inside the contour at 95\% C.L. from the Planck 2015 results. 
}
\label{fig:hp}
\end{figure}
%%%%%%%%%%%%%%%%%%%%%%%%%%%%%%%%%%%%%%%%%%%% 

%%%%%%%%%%%%%%%%%%%%%%%%%%%%%%%%%%%%%%%%%%%%
\begin{figure}[htbp]
\begin{center}
\includegraphics[width=0.45\textwidth,angle=0,scale=1.02]{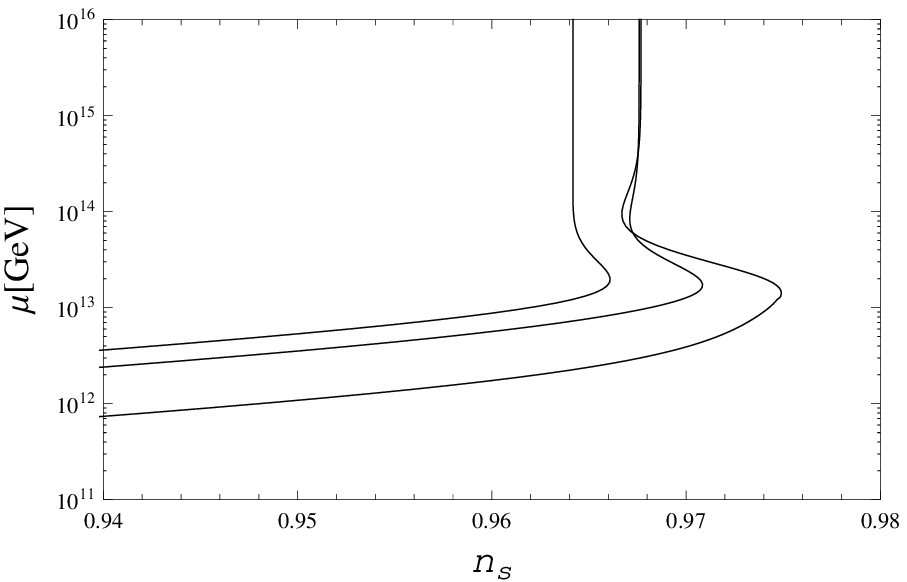} \hspace{0.5cm}
\includegraphics[width=0.45\textwidth,angle=0,scale=1.02]{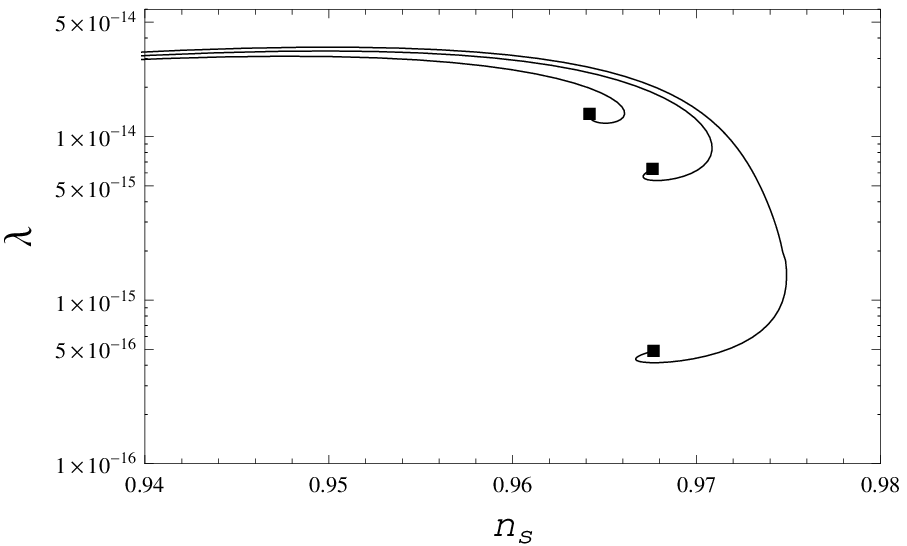}
\end{center}
\caption{
Relations between $n_s$ and $\mu$ (left panel) and between $n_s$ and $\lambda$ (right panel), 
  corresponding to Fig.~\ref{fig:hp}. 
}
\label{fig:hp_mass}
\end{figure}
%%%%%%%%%%%%%%%%%%%%%%%%%%%%%%%%%%%%%%%%%%%% 

We now consider the GB brane-world cosmological effects on the Higgs potential model in the case (ii),  
   and calculate the inflationary predictions for various values of $\mu$ and $v=20$, $30$ and $100$. 
Our results are shown in Fig.~\ref{fig:hp} for $N_0=60$. 
The black squares from bottom to top (top to bottom) in the left panel (the right panel) 
  denotes the inflationary predictions in the standard cosmology limit for $v=20$, $30$ and $100$, respectively. 
As $v$ is raised, the predicted values of $n_s$ and $r$  in the standard cosmology 
  approach those of the quadratic potential model (see the position marked by the black triangle 
  in Fig.~\ref{fig:phi2}). 
As $\mu$ is lowered, the inflationary predictions are deviating from the predictions in the standard cosmology. 
The inflationary predictions run away to the left as we lower $\mu \lesssim 10^{13}$ GeV 
   (see the left panel in Fig.~\ref{fig:hp_mass}), and hence we find lower bounds on 
   $\mu[{\rm GeV}]/10^{12} \geq 5.13$, $3.41$ and $1.03$ for $v=20$, $30$ and $100$, respectively, 
   to be consistent with the Planck 2015 results at 95\% C.L. 
We can see that the brane-world cosmological effect suppresses the tensor-to-scalar ratio 
   for $\mu \lesssim 10^{14}$ GeV. 
Fig.~\ref{fig:hp_mass} shows corresponding results in  ($n_s, \mu$)-plane (left panel) and ($n_s, \lambda$)-plane (right panel). 
In the right panel, the black squares denote the results in the standard cosmology 
  for $v=20$, $30$ and $100$ from top to bottom, respectively.

%%%%%%%%%%%%%%%%%%%%%%%%%%%%%%%%%%%%%%%%%%%%%%%%%%%%%
\subsection{Coleman-Weinberg potential}
%%%%%%%%%%%%%%%%%%%%%%%%%%%%%%%%%%%%%%%%%%%%%%%%%%%%%
Finally, we discuss an inflationary scenario based on a potential with a radiative symmetry breaking~\cite{Shafi_Vilenkin} 
  via the Coleman-Weinberg mechanism~\cite{Coleman_Weinberg}. 
We express the Coleman-Weinberg potential of the form, 
\begin{eqnarray}
 V= \lambda \phi^4 \left[ \ln \left( \frac{\phi}{v}\right)-\frac{1}{4}\right]+\frac{\lambda v^4}{4}, 
\end{eqnarray}
  where $\lambda$ is a dimensionless coupling constant, and $ v$ is the inflaton VEV. 
This potential has a minimum at $\phi=v$ with a vanishing cosmological constant. 
Analysis for the Coleman-Weinberg potential model is analogous to the one of the Higgs potential model. 
Following the same discussion in the previous subsection, 
   we concentrate on the case $\phi_0 < v$ for the initial VEV of the inflaton.

\begin{figure}[htbp]
\begin{center}
\includegraphics[width=0.45\textwidth,angle=0,scale=1.0]{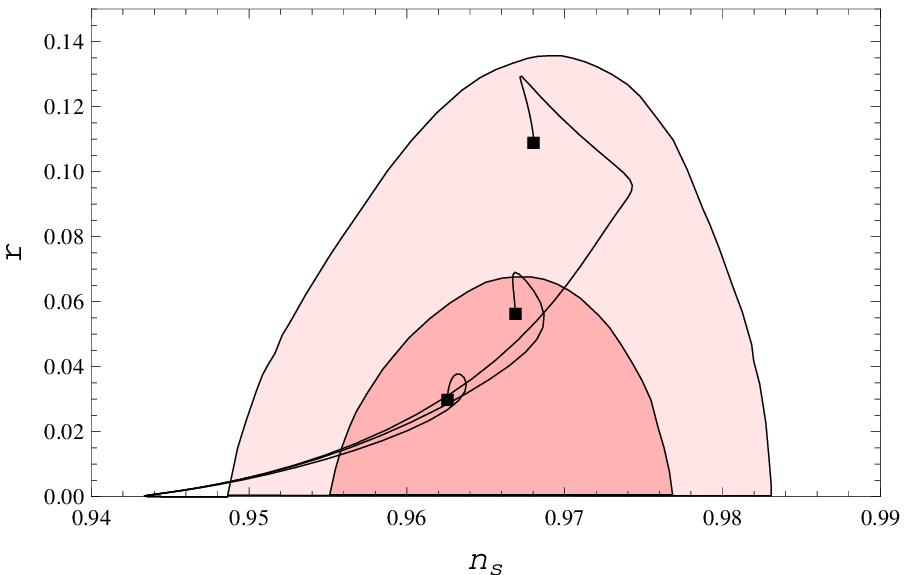} \hspace{0.5cm}
\includegraphics[width=0.45\textwidth,angle=0,scale=1.05]{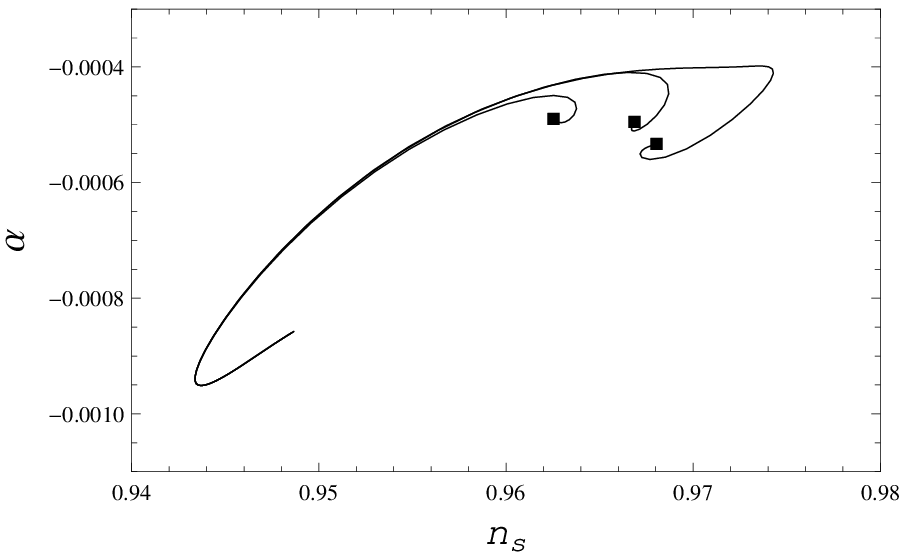}
\end{center}
\caption{
Same as Fig.~\ref{fig:hp} but for the Coleman-Weinberg potential model. 
We have found the lower bounds on  $\mu[{\rm GeV}]/10^{12} \geq 5.01$, $3.27$ and $1.03$ 
  for $v=20$, $30$ and $100$, respectively, for the inflationary predictions 
  to lie inside the contour at 95\% C.L. 
}
\label{fig:CW}
\end{figure}

\begin{figure}[htbp]
\begin{center}
\includegraphics[width=0.45\textwidth,angle=0,scale=1.02]{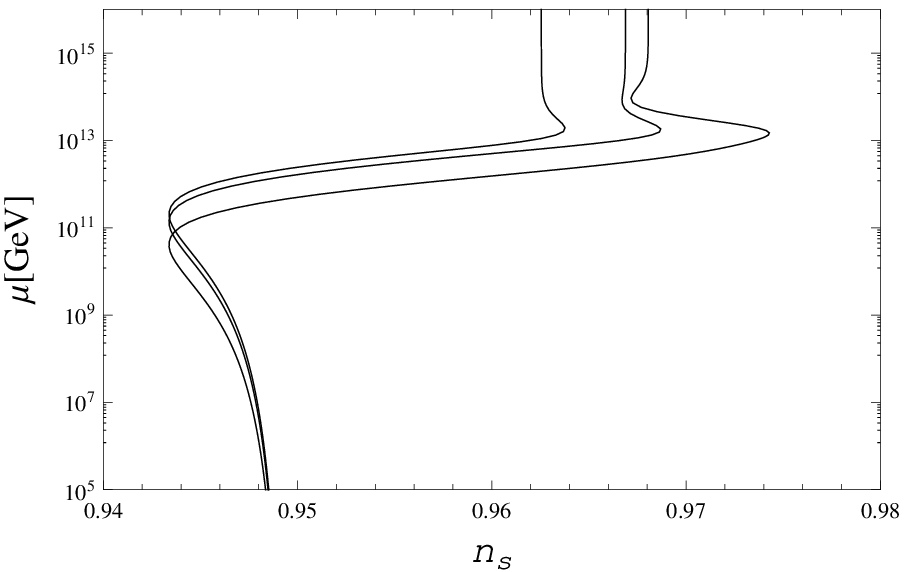} \hspace{0.5cm}
\includegraphics[width=0.45\textwidth,angle=0,scale=1.02]{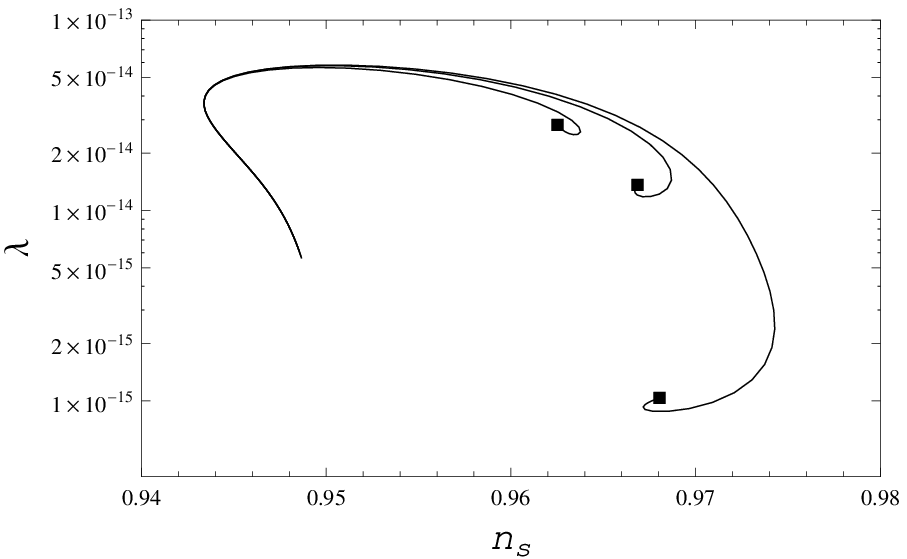}
\end{center}
\caption{
Same as Fig.~\ref{fig:hp_mass} but for the Coleman-Weinberg potential model. 
}
\label{fig:CW_mass}
\end{figure}

We show in Fig.~\ref{fig:CW} the inflationary predictions of the Colman-Weinberg potential model for various values 
 of $v$ and $\mu$ with $N_0=60$. 
The black squares from bottom to top (top to bottom) in the left panel (the right panel) denotes 
  the inflationary predictions in the standard cosmology limit for $v = 20$, $30$ and $100$, respectively. 
As $v$ is raised, the predicted values of $n_s$ and $r$ in the standard cosmology approach 
   those of the quadratic potential model (see the position marked by the black triangle in Fig.~\ref{fig:phi2}). 
As in the Higgs potential model, lowering the $\mu$ value deflects the inflationary predictions 
   from those in the standard cosmology. 
The inflationary predictions run away to the left as we lower $\mu \lesssim 10^{13}$ GeV, 
  but  they return to the right for $\mu \lesssim 10^{10}$ GeV (see the left panel in Fig.~\ref{fig:CW_mass}). 
Similarly to the Higgs potential model, we have find the lower bounds on 
  $\mu[{\rm GeV}]/10^{12} \geq 5.01$, $3.27$ and $1.03$  
  for $v=20$, $30$ and $100$ from the Planck 2015 results.   
Fig.~\ref{fig:CW_mass} shows corresponding results in  ($n_s, \mu$)-plane (left panel) and ($n_s, \lambda$)-plane (right panel).

%%%%%%%%%%%%%%%%%%%%%%%%%%%%%%%%%%%%%%%%%%%%
%%%%%%%%%%%%%%%%%%%%%%%%%%%%%%%%%%%%%%%%%%%% 

%%%%%%%%%%%%%%%%%%%%%%%%%%%%%%%%%%%%%%%%%%%%
%%%%%%%%%%%%%%%%%%%%%%%%%%%%%%%%%%%%%%%%%%%% 

%%%%%%%%%%%%%%%%%%%%%%%%%%%%%%%%%%%%%%%%%%%%%%%%%%%%%%%%
\section{Conclusions}
%%%%%%%%%%%%%%%%%%%%%%%%%%%%%%%%%%%%%%%%%%%%%%%%%%%%%%%%
Observational cosmology is now a precision science, and the cosmological parameters 
   are being very precisely measured. 
Motivated by the Planck 2015 results, we have studied simple inflationary models based on the quadratic, 
  quartic, Higgs and Coleman-Weinberg potentials in the context of the Gauss-Bonnet brane-world cosmology.  
In the presence of the 5th dimensional space, not only the Friedmann equation for our 4-dimensional universe 
  on the brane but also the evolution of scalar and tensor perturbations generated by inflation 
  are modified and as a result, a drastic change of  the inflationary predictions 
  from those in the 4-dimensional standard cosmology can emerge.   
Although the quadratic potential model in the standard cosmology is not favored by the Planck 2015 results, 
  the GB brane-world effect can slightly improve the data fitting for a limited $\mu$ region (for $N_0=60$).   
The quartic potential model is disfavored by the Planck 2015 results, and no improvement  
  has been found by the GB brane-world cosmological effect.  
We have obtained interesting results for the Higgs and Coleman-Weinberg potential models. 
When the GB brane-world cosmological effect is significant, the inflationary predictions of 
  $n_s$ and $r$ are both suppressed and the Planck 2015 results provide us 
  with lower bounds on $\mu$, or equivalently, the 5-dimensional Planck mass, depending on VEVs. 
We have found that the GB brane-world cosmological effect causes  
   a drastic change for the prediction of the spectral index, but a mild change for the tensor-to-scalar ratio. 
Therefore, a precise measurement of the spectral index in the future experiments can narrow 
   an allowed region of the tensor-to-scalar ratio, which can be tested in the future observation of 
   the CMB $B$-mode polarization.

In light of the recent observation of the $B$-mode polarization by the BICEP2 collaboration, 
  simple inflationary models in the context of the Randall-Sundram brane-world cosmology 
  have been investigated in \cite{OO2}.  
It is interesting to compare the results presented in Sec.~3 in this manuscript 
  to those presented in Sec.~2 in \cite{OO2}.    
We can see  that in the RS brane-world cosmology, the brane-world effect causes 
  a drastic change (enhancement) for $r$, rather than $n_s$. 
This is in sharp contrast with the GB brane-world cosmological effect we have found in this paper.

%%%%%%%%%%%%%%%%%%%%%%%%%%%%%%%%%%
\section*{Acknowledgments}
%%%%%%%%%%%%%%%%%%%%%%%%%%%%%%%%%%
We would like to thank Andy Okada for his encouragements during the completion of the present work. 

%%%%%%%%%%%%%%%%%%%%%%%%%%%%%%%%%

\end{document}